\documentclass{aa}
\usepackage{graphicx}
\def\approxgt{\mathrel{\hbox{\rlap{\lower.55ex \hbox {$\sim$}}
        \kern-.3em \raise.4ex \hbox{$>$}}}}
\def\approxlt{\mathrel{\hbox{\rlap{\lower.55ex \hbox {$\sim$}}
        \kern-.3em \raise.4ex \hbox{$<$}}}}
\begin{document}
   \title{The Phoenix galaxy: UGC~4203 re-birth from its ashes?}

   \author{M. Guainazzi
          \inst{1}
          \and
	  G.Matt
          \inst{2}
          \and
          F.Fiore\inst{3}
	  \and
	  G.C.Perola\inst{2}
          }

   \offprints{M.Guainazzi}

   \institute{XMM-Newton Science Operation Center, VILSPA, ESA, Apartado
              50727, E-28080 Madrid, Spain \\
              \email{mguainaz@xmm.vilspa.esa.es}
              \and
              Dipartimento di Fisica, Universit\`a degli Studi Roma Tre
	      Via della Vasca Navale 84, I-00146 Roma, Italy
	      \and
              Osservatorio Astronomico di Roma, Via dell'Osservatorio,
	      I-00144 Monteporzio Catone, Italy
              }

   \date{Received ; accepted }

   \abstract{We report on a dramatic transition between a
	    Compton-thick, reflection-dominated state and a Compton-thin
	    state in the Seyfert~2 galaxy UGC~4203, discovered by 
	    comparing a recent (May 2001) XMM-Newton
	    observation with ASCA observations
	    performed about six years earlier. This transition
 	    can be explained either as a change in the column 
	    density of the absorber, maybe due to moving clouds
            in a clumpy torus, or as the revival of a
	    transient active nucleus, which was in 
	    a phase of very low activity when observed by ASCA.
	    If the latter explanation is correct, spectral
	    transitions of this kind provide observational
	    support to the idea that Compton--thick and 
	    Compton--thin regions coexist in the same source, the former 
	    likely to be identified with the ``torus", the latter with 
	    dust lanes on much larger scales.
   \keywords{X-rays:galaxies --
		galaxies:active --
		galaxies:Seyfert --
		galaxies:individual:UGC~4203 --
		galaxies:nuclei
            }
            }

\authorrunning{Guainazzi et al.}

\titlerunning{The Phoenix galaxy}

   \maketitle
%
%________________________________________________________________

\section{Introduction}

X-ray spectra of Seyfert~2 galaxies are always significantly absorbed
by line--of--sight cold matter
(\cite{awaki91}, \cite{turner97},
\cite{risaliti01}). The X-ray absorber is usually 
associated with the ``torus", responsible - in Seyfert unification
models (\cite{antonucci85}, \cite{antonucci93}) - for the occultation
of the Broad Line Regions (BLR).

This scenario may be oversimplified. Keel
proposed as early as 1980 that the deficiency
of nearly edge-on Seyfert~1 galaxies may
be due to dust obscuration, associated with the
host galaxy. Maiolino
\& Rieke (1995) suggested that intermediate Seyferts
are seen through a 100~pc-scale, Compton--thin absorber coplanar with 
the disk of the host galaxy, whereas only ``strict" Seyfert~2s are
obscured by the inner torus.
Expanding this line of thought, Matt (2000) has recently
proposed an extension of the unification models,
where Compton-thick Seyfert~2s ($N_{\rm H} \approxgt 10^{24}$~cm$^{-2}$)
are seen through
compact, thick matter with a large covering fraction
(to be identified with the ``torus"),
a few tens of parsecs at most far from the nucleus,
whereas Compton-thin Seyfert~2s
($N_{\rm H} \approxlt 10^{24}$~cm$^{-2}$) are absorbed by matter
located at larger distances. The discovery,
with HST high-resolution
images,  that 
dust lanes on scales from tens to hundred parsecs
are common in nearby Seyferts suggests a possible
``optical" counterpart for the Compton--thin, X-ray absorbing matter
(\cite{malkan98};
see also \cite{antonucci90} for similar, earlier 
results on narrow line radio galaxies). Later on, 
Guainazzi et al. (2001) studied
the correlation between the morphology of the
circumnuclear dust
and the properties of X-ray absorption. Their study indicate
that Compton-thin
objects prefer dust-rich nuclear environments, whereas
the presence of Compton-thick material does not appear to be
correlated with the overall dust content. 

Such studies have been hampered so far by the limited number
of Seyfert~2 galaxies for which good quality
X-ray spectra are available. A program has been therefore
started to observe
with the X-ray satellite XMM-Newton (\cite{jansen01})
a large sample of Seyfert~2s, selected
according to their nuclear dust morphology.
We present in this {\it paper} the results
of the observation of the edge-on maser
galaxy (\cite{braatz94}) UGC~4203 (a.k.a. Mrk~1210; $z = 0.013$),
one of the Seyfert~2s which exhibit broad lines
in polarized light (\cite{tran95a}). This observation reveals
an unusual transition between
a Compton-thick and -thin state, when compared with
ASCA observations performed almost six years earlier.

\section{Observations and data reduction}

XMM-Newton observed UGC~4203 on May 5 2001.
The CCD imaging EPIC cameras
(\cite{turner01}. \cite{struder01})
were
operated in Full Frame mode, with the
{\tt THIN} optical filter. UGC~4203
was too faint to be detected by the high resolution
spectroscopy cameras, whose
data will not be discussed in this {\it paper}. Data
were reduced with {\sc SAS v5.1} (\cite{jansen01}), using
the most updated version of the calibration files available at
August 2001. This ensures spectral fitting residuals
within $\pm$10\% for all EPIC cameras, accuracy of the absolute
position reconstruction within 4$\arcsec$, and of the gain
reconstruction $\le$25~eV at 6~keV.
The observation of UGC~4203 was affected by a rather high radiation level.
The p-n was flooded with background events, and spent most
of the observing time in ``counting mode". In this
mode any information about the position or energy of the incoming
photons is  lost. For this reason, only
MOS data will be discussed in this {\it paper}. After data
screening, the exposure time is about 7.7~ks. 

In this {\it paper}:
errors refer to the 90\% confidence level for one interesting
parameter; energies are quoted in the source rest frame,
unless otherwise specified.

\section{Spectral analysis}

\subsection{The XMM-Newton observation}

In the MOS
cameras, a strong source
(MOS~2 count rate in the 0.3--10~keV band
of $0.220 \pm 0.006$~s$^{-1}$)
is detected. Its centroid position
($\alpha_{2000}=08^h04^m05^s.8$;
$\delta_{2000}=+05^h06^m50^s$) coincides within 2$\arcsec$ with
the optical nucleus of UGC~4203.
The MOS image
does not show any evidence of extended emission beyond the instrumental 
PSF (6$\arcsec$ correspond to 100~kpc at the UGC~4203 distance). No
flux variation is observed during the observation. We will
therefore focus on the time-averaged spectrum only in this {\it paper}.
X-ray events corresponding to patterns 0 to 12 were used.
Source spectra were extracted from a circular region around the
image centroid, with a radius of 22$\arcsec$.
Spectra were rebinned in order to have at least
50 counts for each spectral bin, to allow a proper use of the $\chi^2$ test,
and to oversample
the instrumental energy resolution by a factor not higher than 3.
Background spectra were extracted from several
regions of the central chip, and merged together to increase
statistics, once verified that the results were not significantly
dependent on the chosen extraction region. A fit
to the MOS spectra with a simple, absorbed power-law
in the 3--10~keV energy range
is acceptable ($\chi^2 = 57.0/52$~dof).
Excess counts at an observed energy of about
6.3~keV can be interpreted as an emission line,
as explained in more details later in this Section.
The photoelectric column density is
$N_{\rm H} = 1.8 \pm^{0.4}_{0.3} \times 10^{23}$~cm$^{-2}$,
whereas the spectral index, albeit only loosely constrained,
is not dissimilar from that typically observed in Seyfert~2s
($\Gamma = 1.6 \pm^{0.3}_{0.4}$).

\begin{figure}
   \centering
   \includegraphics[angle=-90,width=8cm]{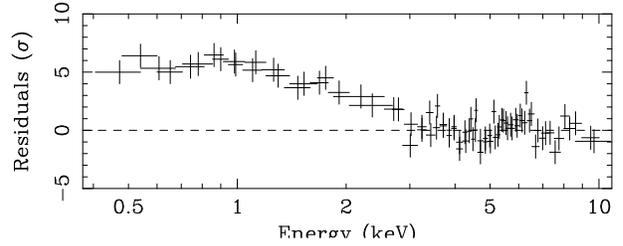}
      \caption{Residuals in units of standard deviations
	       when the best-fit
	       photoelectric absorbed power-law model,
	       applied in the 3--10~keV band, is
	       extrapolated redwards.
              }
         \label{fig1}
   \end{figure}

The extrapolation
of the best-fit model in the 3--10~keV band lays, however, well below
the data (see Fig.~\ref{fig1}).
%-------------------------------------------------------------
%%-------------------------------------------------------------
We have therefore added to the hard X-ray
absorbed power-law a second unabsorbed power-law,
to approximately describe
the scattering of the primary nuclear emission by
a ``warm mirror".
Photoelectric absorption with a column density
equal to that of our own Galaxy
($N_{\rm H,Gal} = 3.7 \times 10^{20}$~cm$^{-2}$; \cite{dickey90})
has been added to all  models hereinafter.
If self-absorption effects are
neglected, the scattered soft component has the
same spectral index of the primary emission;
the data do not actually require different indices. The scattering fraction
is $1.3 \pm^{0.3}_{0.7} \%$. Other descriptions of the soft
excess, like {\it e.g.} thermal emission from
an optically thin collisionally ionized plasma
which may be produced in regions of intense star formation,
are, however, also consistent with the data.
The results on the hard X-ray spectrum are not
significantly dependent on the soft excess modeling.

\begin{figure}
   \centering
   \includegraphics[angle=-90,width=8cm]{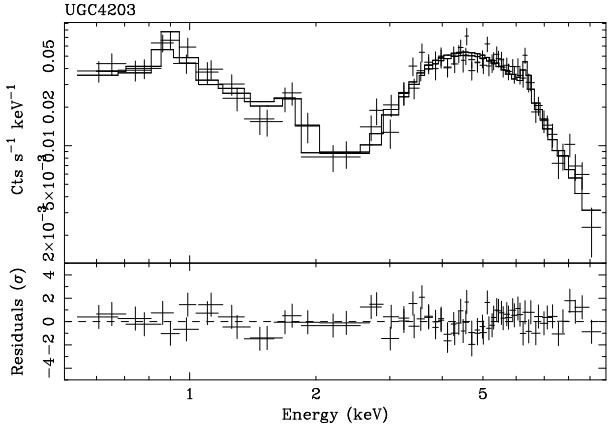}
      \caption{UGC~4203 MOS spectra ({\it upper panel}) and
	       residuals in units of standard deviations
	       ({\it lower panel}) when the best-fit
	       scattering scenario model is applied.
              }
         \label{fig2}
   \end{figure}

The best-fit parameters and results are summarized in Table~\ref{tab1}.
%-------------------------------------------------------------
\begin{table*}
\begin{center}
\begin{tabular}{cccccc} \hline \hline
$\Gamma$ & $N_{\rm H}$ & $E_{\rm c}$ & $I_{\rm c}$ & $EW$ &  $\chi^2/$~dof \\
& ($10^{23}$~cm$^{-2}$) & (keV) & ($10^{-5}$~cm$^{-2}$~s$^{-1}$) & (eV) & \\ \hline
$2.0 \pm 0.2$ & $2.14 \pm^{0.20}_{0.16}$ & $0.93 \pm^{0.02}_{0.05}$ & $1.7 \pm^{0.4}_{0.9}$ & $130$$^{\dag}$ & 65.1/70 \\
& & $1.79 \pm^{0.06}_{0.09}$ & $4 \pm^4_3$ & $110$$^{\dag}$ &  \\
& & $6.41 \pm 0.06$ & $2.9 \pm^{2.3}_{1.3}$ & $130$$^{\ddag}$ & \\ \hline \hline
\end{tabular}

\noindent
$^{\dag}$against the scattered continuum; $^{\ddag}$against the primary continuum

\end{center}

\caption{Best-fit parameters and results of the XMM-Newton observation of UGC~4203. Model scenarios are explained in text}
\label{tab1}
\end{table*}
%-------------------------------------------------------------
Three narrow ({\it i.e.}: intrinsic width $\sigma = 0$)
emission lines are also required
to get an acceptable fit ($\chi^2 = 65.1/70$~dof).
One of them ($E_{\rm c} \simeq 6.41$~keV, $\Delta \chi^2 = 14.1$,
corresponding to a confidence level $\simeq$99.9\%) is
consistent with K$_{\alpha}$ fluorescent emission from
neutral iron. The other two lines have
$E_{\rm c} \simeq 0.93$~keV and
$E_{\rm c} \simeq 1.79$~keV, respectively. The
corresponding improvement
in the quality of the fit is $\Delta \chi^2 = 14.1$
and $\Delta \chi^2 = 4.2$, respectively, the latter
corresponding to a confidence level of 88.8\% only.
Their centroid energies are consistent with K$_{\alpha}$
fluorescent transitions of H-like neon and silicon,
probably due to radiative recombination and resonant scattering 
in a photoionized plasma (\cite{bianchi00},
\cite{sako00}, \cite{sambruna01}). All the detected
lines are unresolved, even if the upper limits to their widths
are very large. Spectrum and residuals for the best-fit model
are shown in Fig.~\ref{fig2}.
%-------------------------------------------------------------
%%-------------------------------------------------------------

The 0.5-2.0~keV (2.0-10~keV) flux measured by XMM-Newton
is $2.6 \times 10^{-13}$~erg~cm$^{-2}$~s$^{-1}$
($9.7 \times 10^{-12}$~erg~cm$^{-2}$~s$^{-1}$), 
corresponding to a rest-frame, absorption-corrected
2--10~keV luminosity
of $1.9 \times 10^{43}$~erg~s$^{-1}$ for the
transmitted nuclear emission.

\subsection{ASCA observation}

\begin{figure}
   \centering
   \includegraphics[angle=-90,width=8cm]{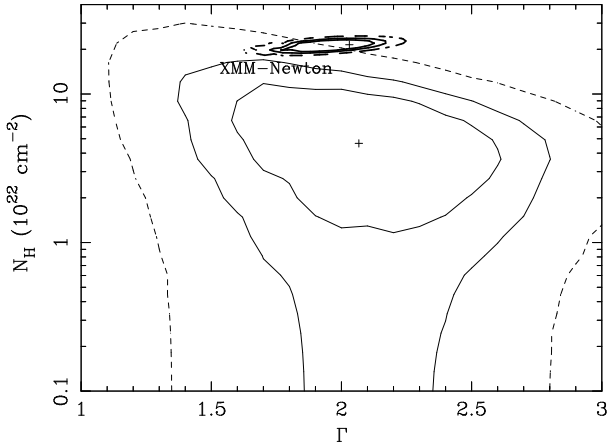}
      \caption{Iso-$\chi^2$ contour plot for the
		intrinsic power-law spectral index
		versus the Compton-thin column density
		in the ASCA and XMM-Newton (small
		contours in {\it bold}) observation.
		In the former, the Compton-thin absorber
		covers the reflection-component only.
		Contours are at 68\%, 90\% ({\it
		solid lines}) and 99\% ({\it solid line})
		confidence
		level for two interesting parameters
              }
         \label{fig6}
   \end{figure}

ASCA observed UGC~4203 twice during autumn 1995,
approximately
three weeks apart (October 18 and November 12). 
An analysis of the ASCA data is presented by
Awaki et al. (2000).
We have
combined the two observations and
reanalyzed the data, obtaining
basically their same results.
The hard X-ray spectrum during the ASCA observation
was significantly fainter than during the XMM-Newton one.
The fluxes measured by ASCA are
$2.3 \times 10^{-13}$~erg~cm$^{-2}$~s$^{-1}$,
and $2.1 \times 10^{-12}$~erg~cm$^{-2}$~s$^{-1}$
in the 0.5--2~keV and 2--10~keV energy band, respectively.
We have first tried to fit the spectra with
the same model adopted to fit the XMM-Newton spectrum
({\it transmission scenario} in Table~\ref{tab4}).
Although the fit is acceptable ($\chi^2_{\nu} = 0.89$)
there is no
obvious explanation for the huge ($EW \simeq 1.2$~keV)
iron emission line (\cite{leahy93}, Matt 2002). This feature is instead
naturally explained in the reflection-dominated
scenario, appropriate for nuclei either absorbed by Compton--thick matter 
(\cite{matt96}) or switched--off (\cite{guainazzi98}), 
where emission lines are observed against the
reflection continuum only (\cite{matt96}).
We have therefore substituted the
absorbed hard X-ray power-law
with an unabsorbed Compton-reflection component
(model {\tt pexrav} in {\sc Xspec}, \cite{magdziarz95};
{\it reflection-dominated
scenario} in Table~\ref{tab4}).
%-------------------------------------------------------------
\begin{table*}
\begin{center}
\begin{tabular}{lccccccc} \hline \hline
Scenario & $\Gamma$ & $N_{\rm H}$ & $f_{\rm s}$ & $E_{\rm c}$ & $I_{\rm c}$ & $EW$ & $\chi^2/$~dof \\
& & ($10^{23}$~cm$^{-2}$) & (\%) & (keV) & ($10^{-5}$~cm$^{-2}$~s$^{-1}$) & (eV) & \\ \hline
Transmission & $1.8 \pm^{0.4}_{0.5}$ & $2.6 \pm^{1.7}_{0.8}$ & $10 \pm^7_{10}$ & $6.44 \pm^{0.07}_{0.06}$ & $4.9 \pm^{1.4}_{1.6}$ & 1160 & 90.0/101 \\
Reflection-dominated$^a$ & $2.0 \pm^{0.5}_{0.6}$ & ... & ... & $6.43 \pm^{0.09}_{0.06}$ & $3.1 \pm^{1.1}_{0.9}$ & 1010 & 93.6/102 \\
Reflection-dominated$^b$ & $2.0 \pm 0.5$ & $0.5 \pm^{0.7}_{0.4}$ & ... & $6.44 \pm^{0.07}_{0.06}$ & $3.1 \pm 1.0$ & 880 & 87.8/101 \\ \hline \hline
\end{tabular}
\end{center}

\noindent
$^a$the reflection component is photoelectrically absorbed by $N_{\rm H,Gal}$ only

\noindent
$^b$the reflection component is photoelectrically absorbed by a screen of column density $N_{\rm H}$

\caption{Best-fit parameters and results of the ASCA observation of UGC~4203. Model scenarios are explained in text. $N_{\rm H}$ is the absorption of the nuclear
primary power-law, when the transmission scenario is employed to model the
hard X-ray emission}
\label{tab4}
\end{table*}
%-------------------------------------------------------------
We fixed the abundances to their
solar values, and assumed reflection from an edge-on plane--parallel slab.
This model describes the data equally well ($\chi^2_{\nu} = 0.90$), but it
is to be preferred on physical grounds. Even if the fit is already
satisfactory, we have later included in the model an additional
photoelectric absorption covering the Compton-reflection
component, to verify whether the latter is also obscured
by a Compton-thin absorber similar to the one found in
the XMM-Newton observation. The the quality of the fit slightly improves 
($\Delta \chi^2=$5.8,
corresponding to the 98.8\% confidence level according
to the F-test).
This remains true also if
the inclination angle and the abundances are left free to vary
in the fit. 
The best-fit column density value is four times lower than,
but still consistent at the 3$\sigma$ level (see Fig.~\ref{fig6})
with that derived from the XMM-Newton spectrum.
%-------------------------------------------------------------
%%-------------------------------------------------------------

The ASCA spectra do not provide significant constraints
on the iron line profile or on
the nature of the soft X-ray emission.
Despite the remarkable flux change
in the 2--10~keV band between the
XMM-Newton and the ASCA observations, the intensity
of the iron line remained basically unchanged.

\section{Discussion}

\begin{figure}
   \centering
   \includegraphics[width=8cm]{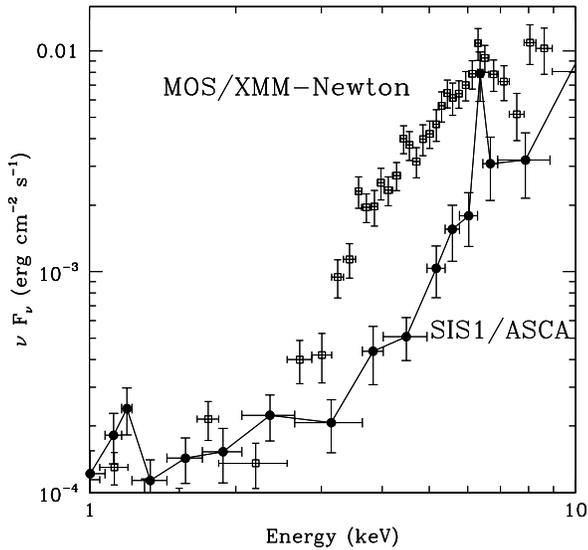}
      \caption{X-ray spectral energy distribution of UGC~4203
	       during the ASCA (connected {\it filled circles}) and
	       the XMM-Newton ({\it empty squares}) observations
              }
         \label{fig4}
   \end{figure}

The two X-ray observations of UGC~4203 described in
this {\it paper} caught the source in two
remarkably different spectral states.
The ASCA spectrum is most-likely
dominated by Compton-reflection.
We assume in what follows that this
component is due to reflection from the far inner
side of Compton-thick matter surrounding
the nucleus. We are therefore implicitly assuming an
axisymmetric distribution of this
matter, as in the torus envisaged in the
standard unification scenarios (\cite {antonucci85},
\cite{antonucci93}). In
the XMM-Newton observation of UGC~4203, however,
the AGN was shining again (albeit through a Compton-thin
absorber with $N_{\rm H} \simeq 2 \times 10^{23}$~cm$^{-2}$).
In Fig.~\ref{fig4} we compare the UGC~4203
%-------------------------------------------------------------
%%-------------------------------------------------------------
X-ray spectral
energy distributions measured by ASCA and XMM-Newton.
Despite the 2-10~keV flux difference, the
intensities of the iron line in the two observations are
consistent with each other. It is interesting to note
that comparable fluxes are measured also in the
soft X-rays, which are likely to
be dominated by reprocessing or diffuse emission.
The lack of information about the history of the
activity in UGC~4203 prevents us from deriving
quantitative constraints on the location of the
matter, which emits the bulk of the
radiation at $\approxlt$2~keV.

\begin{figure}
   \centering
   \includegraphics[width=8cm]{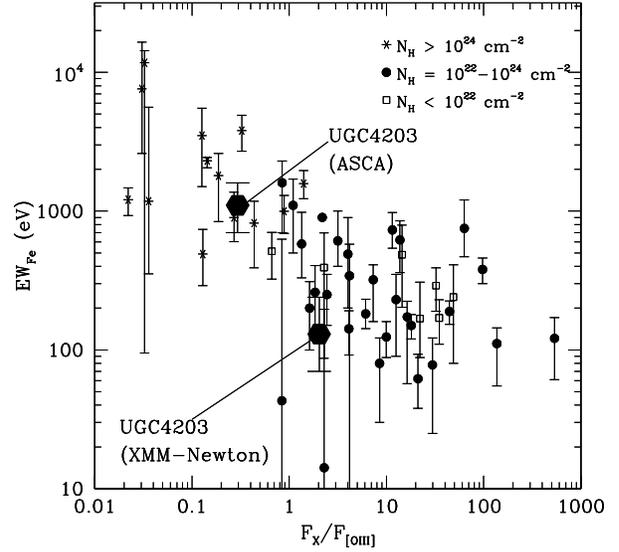}
      \caption{``Bassani" diagnostic plot: K$_{\alpha}$
		fluorescent iron 
	       line EW against the ration between the
	       2--10~keV and the Balmer-decrement
	       corrected O{\sc iii} flux. The position
	     	of UGC~4203 in the ASCA and XMM-Newton
		observation is indicated by the big
		{\it filled exagons}
              }
         \label{fig5}
   \end{figure}

In Fig.~\ref{fig5},
the transition of UGC~4203 between a ``Compton-thick" and a ``Compton-thin"
state is
%-------------------------------------------------------------
%%-------------------------------------------------------------
put in the context of the diagnostic diagram of Bassani et al. (1999).
The iron line EW is plotted against the ratio between the
2--10~keV and the O[{\sc III}]
flux corrected for the Balmer-decrement (taken from \cite{ho97}).
The path followed by UGC~4203 in this diagram is
that expected from the  above transition.
Although variations of the X-ray
absorber in Compton-thin objects (rarely larger than
a factor of a few)
are quite common (\cite{risaliti01}), there are only
two other
convincing cases where a transition between a Compton-thin
and -thick state can be claimed: NGC~1365; (\cite{iyomoto97},
\cite{risaliti00}), and NGC~6300 (\cite{guainazzi01b}),
out
of about 30 objects for which at least two historical measurements
of $N_{\rm H}$ are available.

The transition of UGC~4203 can be due to either: 1)
a change in the properties of the
absorber(s); 2) or a ``switching--off" of the nucleus 
during the ASCA observation, leaving the
Compton-reflection as the only ``echo" of its previous
higher activity level.

The timescale, on which the transition
occurred, rules out an explanation in terms
of ``disappearance" of a single, homogeneous Compton-thick
axis-symmetric system, covering a steady-state
active nucleus.
The covering fraction of the torus must be
$\sim 2 \pi$ to account for the residual flux observed
by ASCA.
Large torus covering fractions are also suggested
by statistical arguments (\cite{maiolino95};
\cite{risaliti00}).
If  the observed transition is related
to the line-of-sight crossing time of a single
spherical cloud in Keplerian motion around a
$M_8 \equiv M/10^8 M_{\odot}$ black hole, the
distance from the central engine is
$\sim 8 \times 10^{16} M_8^{1/3}$~cm.
A rather unlikely
geometry of the torus would be necessary to
explain the transition in this scenario.

However, an explanation along the same line of thought
is viable, if the Compton-thick matter is clumpy.
The crossing-time of a Keplerian cloud covering
a region of size $r_{10} = r/10 R_{\rm S}$
around the black hole is:
$$
t_{\rm c} \simeq 0.13 M_8^{1/2} d^{1/2} r_{10} \ \hbox{years}
$$
where $d$ is the cloud distance from the
nucleus in parsecs.

Alternatively, ASCA may  have observed  the
AGN in UGC~4203 in a dim state.
Compton-reflection would be therefore
only the ``echo" of its previous activity, later revived, and
discovered by the XMM-Newton observation
(the ``Phoenix galaxy" effect).
UGC~4203 would in this case join, {\it e.g.},
NGC~4051 (\cite{guainazzi98}; \cite{uttley99})
and NGC~2992 (\cite{gilli00}),
whose dimming and subsequent recovery on
timescales of the order of a fraction of
year to a few years have been already 
monitored. In both cases the fainter
states correspond to flat X-ray spectra with
strong K$_{\alpha}$ fluorescent
iron lines, interpreted as the emerging of
reflection from Compton-thick matter at distances
$\approxgt 0.1$~pc.

The implications of the latter explanation would be
far-reaching. In the framework of the Seyfert
unification scenario, X-ray reflection-dominated
Seyfert~2 galaxies are strictly identified with
Compton-thick objects. Recent hard X-ray studies
(\cite{salvati97}, \cite{maiolino98}, \cite{risaliti00})
suggest that they may constitute about half of the
Seyfert~2 population, at least in the local
Universe. If, on the other hand, reflection-dominated states
are due to AGN low-activity phases, their rate of occurrence may
be telling us more on the duty-cycle of the
AGN phenomenon, rather than on the geometry 
of the nuclear environment. 

It is admittedly difficult to 
discriminate between the two above scenarios,
given the paucity of information on the
X-ray history of UGC~4203. In NGC~6300 (\cite{guainazzi01b}),
a similar transition from a reflection- to a transmission-
dominated status was observed. However, a remarkably
strong residual Compton-reflection component was clearly present
in the latter state, a factor of 4 higher than typically
observed in unabsorbed Seyfert galaxies. Since
in the latter the reflector covers probably a $\approx \pi$
solid angle to the black hole, the most straightforward
explanation for this finding is to
invoke an intrinsic change of the AGN primary emission,
probably larger than an order of magnitude. NGC~6300
is the best demonstration to date that the
reflection- to transmission-dominated
transitions are indeed related to variation
of the AGN intrinsic luminosity.
Unfortunately, the limited energy band
of the EPIC cameras prevents
us from verifying this hypothesis in the UGC~4203 case.

\subsection{Where is/are the absorber/s?}

Independently of which of the above explanations
is correct, the
results presented in this {\it paper}
have implications on the nature and/or location
of the matter responsible for the X-ray
photoelectric absorption in Seyfert~2
galaxies.

A ``clumpy torus" may
explain the observed transition in UGC~4203.
We may be observing UGC~4203 through the
unstable or uneven torus rim, causing
clouds of different column densities to appear
along the line-of-sight to the nucleus due
to local instabilities or AGN-driven
clouds evaporation (\cite{pier95}).
Clumpy torii may be thermally and dynamically
supported by the AGN radiation field, as shown by
Pier \& Krolik (1992). Magnetic fields may play 
a role  as well. Although
more theoretical work and fully self-consistent
calculations are needed, this seems
a rather promising possibility to explain
our results.

If, on the other hand, the transition 
is due to a changed AGN activity, it follows that
the Compton-thin absorbing matter may
be different than the torus, whose presence 
in UGC~4203 is implied by the
reflection-dominated, X-ray dim state observed by ASCA.
A possible location of this additional absorber
was suggested by 
Matt (2000). In his extension of
the traditional unification scenario, the
compact, dusty torus is still responsible for the
Compton-thick X-ray nuclear absorption only,
whereas the Compton-thin
absorbers are located at much larger distances.
If this scenario applies to UGC~4203 as well, the
Compton spectrum in the reflection-dominated state should be
absorbed by Compton-thin matter. The quality of the 
ASCA spectrum is unfortunately not good enough to give a
definitive answer to this question.
However, an additional
Compton-thin absorber is required at
the 98.8\% level in the ASCA reflection-dominated state.

Alternatively, BLR clouds are expected to have
a column density of $N_{\rm H} \approx 10^{23}$~cm$^{-2}$ as well.
However, the possibility that they are responsible for the
X-ray Compton-thin absorption
is ruled out by the low covering fraction
of the BLR as estimated from the lack of Ly-edge
cutoff or carbon dip in the UV spectrum of type~1
AGN (\cite{maiolino01}) and the lack
of silicate absorption feature in the ISO spectra of
AGN (\cite{clavel00}).

\section{Summary}

Comparing an autumn 1995 ASCA with a summer 2001 XMM-Newton
observation of the Seyfert~2 galaxy UGC~4203 (Mkn~1210),
we discovered a rare transition from a
Compton reflection- to a transmission-dominated state.
This transition can be explained by one of the following
scenarios:

\begin{itemize}

\item[-] the ``disappearance" of a Compton-thick cloud,
covering a steady-state active nucleus at the time
of the ASCA observation. If this explanation is correct,
the outcomes presented in this {\it paper} suggest
that the compact
molecular torus, encompassing the nuclear environment in
the Seyfert unification scenarios, has actually
a clumpy or patchy structure

\item[-] a transition in the activity level of the
UGC~4203 AGN. The XMM-Newton would have unveiled
its revival after the ``silent" phase
caught by ASCA (the ``Phoenix galaxy"
effect)

\end{itemize}

The latter explanation would imply that, at least in
this object [and probably in the objects
where this kind of transitions occurs (\cite{guainazzi01b})]
the Compton-thin and Compton-thick absorbers are 
different.  This would provide support to models where
the standard ``torus" intercepts the line of
sight to the nucleus
in Compton-thick objects only (\cite{matt00}), whereas Compton-thin
absorbers are located at larger scales, maybe
associated with the host galaxy rather then with the
nuclear environment.

\begin{acknowledgements}

Constructive comments from an anonymous referee, which
greatly improved the quality of the presentation, are
gratefully acknowledged. The authors
acknowledge as well helpful discussions with J.Krolik
on the radiation-supported torii models, and
stimulating historical remarks by R.R.J.Antonucci.
FF, GCP and GM
acknowledge financial support from ASI and
the Italian Ministry of Research, under grant
{\sc Cofin-00-02-36}.

\end{acknowledgements}

\end{document}